\documentclass{svproc}
\usepackage{url}
\usepackage{graphicx}  
\usepackage{caption}   
\usepackage{amsmath}
\usepackage{array}

\usepackage{cite}
\usepackage{subcaption}
\usepackage{multirow}
\usepackage{makecell}    
\usepackage{subcaption}
\usepackage{xcolor}
\usepackage{booktabs} 
\usepackage{colortbl} 
\usepackage{soul}
\usepackage{hyperref}
\usepackage{tikz}
\newcommand{\darklabel}[1]{%
  \tikz[baseline=(char.base)]{
    \node[shape=circle, fill=gray!60, text=black, inner sep=1pt] (char) {\small #1};
  }%
}

\begin{document}
\mainmatter              
\title{Weak Links in LinkedIn: Enhancing Fake Profile Detection in the Age of LLMs}
%
%
\author{Apoorva Gulati\inst{1}, Rajesh Kumar\inst{2}, Vinti Agarwal\inst{1}, \and Aditya Sharma\inst{1}}
\institute{BITS Pilani, India \\
\email{\{f20210934, vinti.agarwal, p20200470\}@pilani.bits-pilani.ac.in} 
\and
Bucknell University, USA, \email{rajesh.kumar@bucknell.edu}}



\maketitle              

\begin{abstract}
Large Language Models (LLMs) have made it easier to create realistic fake profiles on platforms like LinkedIn. This poses a significant risk for text-based fake profile detectors. In this study, we evaluate the robustness of existing detectors against LLM-generated profiles. While highly effective in detecting manually created fake profiles (False Accept Rate: $6-7$\%), the existing detectors fail to identify GPT-generated profiles (False Accept Rate: $42-52$\%). We propose GPT-assisted adversarial training as a countermeasure, restoring the False Accept Rate to between $1-7$\% without impacting the False Reject Rates ($0.5-2$\%). Ablation studies revealed that detectors trained on combined numerical and textual embeddings exhibit the highest robustness, followed by those using numerical-only embeddings, and lastly those using textual-only embeddings. Complementary analysis on the ability of prompt-based GPT-4Turbo and human evaluators affirms the need for robust automated detectors such as the one proposed in this study. 

\end{abstract}


\section{Introduction}
Online professional networks, such as LinkedIn, play a crucial role in professional interactions, hosting over $1.15$ billion active users and generating significant economic activity \cite{demandsage2025}. However, such platforms face growing threats from fake profiles used for phishing, misinformation, and recruitment fraud \cite{cao2012aiding, adikari2020identifyingfakeprofileslinkedin}. Recent advances in Large Language Models (LLMs), particularly GPT-3.5 and GPT-4, have simplified the creation of highly realistic fake profiles, posing a significant threat to the existing detectors \cite{Ayoobi_2023,strandell2024linkedin}. Between $2021$ and $2022$, the number of fake profiles on LinkedIn nearly doubled \cite{strandell2024linkedin}. Prompt-based evaluations of humans and GPT-4 achieved modest detection accuracy (F1 Human: $59$\%, F1 GPT-zero shot: $71$\%, F1 GPT-few shot: $86$\%, see Section \ref{sec:human_benchmark_results}). Existing detection approaches, such as Section and Subsection Tag Embeddings (SSTE) proposed in \cite{Ayoobi_2023}, perform well (F1$\sim$$96$\%) against manually created fake profiles but fail sharply (F1$\sim$$68$\%) against LLM-generated profiles.

To address these challenges systematically, we pose and address \darklabel{$q_1$} How vulnerable are current detection methods to profiles generated by advanced LLMs? \darklabel{$q_2$} Can adversarial training with LLM-generated profiles enhance detection robustness? \darklabel{$q_3$} How does the effectiveness of our proposed detection methods compare to human evaluators and GPT-4? 

Our primary contributions are as follows. \darklabel{1} We implement a series of robust fake profile detection systems using textual, numerical, and fused features, incorporating Section Tag Embeddings (STE), Section and Sub‑Section Tag Embeddings (SSTE), along with PCA-based dimensionality reduction. The best setup outperformed prior methods on genuine and manual fake profiles ~\cite{gulati2024linkedin}. \darklabel{2} We augment the existing dataset with 600 fake profiles that we generated using GPT-4-Turbo with carefully crafted prompts. These synthetic profiles closely mimic legitimate users, as verified by similarity metrics, and used for creating attack vectors. \darklabel{3} We demonstrate that existing detectors fail against LLM-generated profiles (FAR: $42-52$\%) and proposed GPT‑assisted adversarial training, which restored FAR to $1-7$\% without compromising the legitimate user classification. \darklabel{4} We also benchmark detection capabilities of human annotators and GPT‑4, confirming the need for ML-based automated detectors. \darklabel{5} Finally, we conducted ablation studies revealing text embeddings alone are fragile under LLM attack, whereas numerical profile features remain sturdier; their fusion yields the most robust detector.

The remainder of this paper is structured as follows: Section~\ref{sec:related-work} reviews related literature, Section~\ref{sec:materials-methods} describes materials and methods, Section~\ref{sec:results} reports and discusses results, Section~\ref{sec:future_dir} limitations and future research directions, and Section~\ref{sec:conclusion} concludes. 
 
\section{Related work}
\label{sec:related-work}
Research on fake profile detection spans numerical, graph-based, behavioral, and textual methods. Early efforts used correlation-based analysis of profile attributes. For instance, Adikari et al. \cite{adikari2020identifyingfakeprofileslinkedin} achieved $87.34$\% accuracy on LinkedIn profiles; however, their approach relied on historical data and assumed attribute consistency, which are limitations when handling cold-start accounts. Graph-based models, such as SybilBelief \cite{Gong2014}, SybilEdge \cite{breuer2020friendfauxgraphbasedearly}, and SybilFlyover \cite{Li2022}, leverage network topology and user connectivity, often achieving AUCs above $0.9$. However, they require relational metadata (e.g., connections, followers), \textit{limiting their applicability for newly created or minimally active profiles}. Early stylometric techniques relied on N-grams and writing patterns \cite{Barbon2016}. The LLM-assited fake profile detection problem is similar to LLM-assisted cheating detection \cite{kundu2024keystroke,roh2025llm}. Recent keystroke dynamics-based approaches \cite{Kuruvilla2024, Bhattasali2021} achieve a detection accuracy close to $95$\%. Using activity-based features such as post frequency and follower-following ratios, Alnagi et al. \cite{Alnagi2024} employed XGBoost with SHAP-based interpretability, achieving $94$\% precision on Instagram and $91$\% on Twitter. 

Ayoobi et al. \cite{Ayoobi_2023} proposed Section and Subsection Tag Embeddings (SSTE), reaching $95$\% accuracy on genuine and manually crafted fake LinkedIn profiles. However, their performance drops sharply (to $71–76$\% accuracy) against GPT-generated profiles, revealing a growing vulnerability. Our work differs from Ayoobi et al. \cite{Ayoobi_2023} and focuses on (1) investigating the robustness of baseline detectors (trained on genuine and manually created fake profiles) against profiles generated by GPT3.5 and GPT4Turbo, (2) assessing the power of GPT3.5, GPT4Turbo, and GPT3.5+GPT4Turbo-assisted adversarial training, and (3) evaluating the performance of Human and GPT-based evaluators systematically.  
 
\section{Materials and methods}
\label{sec:materials-methods}

\subsection{Augmenting existing dataset}
We augment the dataset presented by Ayoobi et al.~\cite{Ayoobi_2023} by adding $600$ GPT-4-generated profiles (GPT4Ps), resulting in $4,200$ profiles: $1,800$ legitimate LinkedIn profiles (LLPs), $600$ manually crafted fakes (FLPs), $1,200$ GPT-3.5-generated profiles (GPT3.5Ps), and $600$ GPT4Ps. GPT3.5Ps were created using zero-shot prompting from both LLP and FLP templates. GPT4Ps were created using few-shot prompting with curated LLP exemplars and GPT-4 Turbo. All profiles adhere to LinkedIn's structure, featuring fields for name, location, education, work history, skills, recommendations, and summary. Prompts and similarity-based quality validation details are provided on a dedicated webpage \cite{paperwebsite}. 

\subsection{Feature extraction}
\label{sec:feature}
Each profile was cleaned and parsed: malformed records were corrected, composite entries split, and essential fields (Name, Experience, Education, Location) verified. We extracted $17$ \ul{numerical features} that capture profile structure, including the count of jobs, education entries, skills, recommendations, followers, and connections. 

For \ul{textual features}, we tested six encoders: BERT~\cite{devlin-etal-2019-bert}, RoBERTa~\cite{liu2019roberta}, DeBERTa-v3~\cite{he2021deberta}, ModernBERT~\cite{warner2024modernbert}, Flair~\cite{akbik2018contextual}, and GloVe~\cite{pennington2014glove}. PCA was applied to reduce embeddings to $150$ dimensions from $786$, improving robustness by eliminating low-ranked components~\cite{bhagoji2017enhancingrobustnessmachinelearning, demontispca}. We selected RoBERTa, ModernBERT, DeBERTa, and Flair for further analysis. 

We simplify the original Section and Subsection Tag Embeddings (SSTE)~\cite{Ayoobi_2023} to Section Tag Embeddings (STE), aggregating each section’s text (e.g., Education) as a single unit, and computed as \( F = \frac{1}{N} \sum_{j=1}^{N} \left(E_j - \text{Em}(\text{Tag}_j)\right) \), where \(E_j\) is the embedding of the \(j\)-th section's text, \(\text{Em}(\text{Tag}_j)\) is the embedding of its tag, and \(N\) is the total number of sections.  Text-embeddings were concatenated with $17$ normalized numerical features, yielding a $167$-dimensional profile vector.

\subsection{Choice of classifiers and hyperparameter tuning}
We used six classifiers: Logistic Regression, Random Forest, SVM, KNN, XGBoost~\cite{chen2016xgboost}, and CatBoost~\cite{prokhorenkova2018catboost}, reflecting standard choices in prior work~\cite{Ayoobi_2023}. XGBoost and CatBoost consistently performed best and were selected for full evaluation.

Hyperparameters were tuned using Bayesian Optimization (BO)~\cite{snoek2012practicalbayesianoptimizationmachine, bergstra2011algorithms} and Genetic Algorithms (GA)~\cite{young2015optimizing}. BO used $30$ trials on a validation split, followed by $20$ trials with five-fold cross-validation. GA used $50$ individuals over three generations, followed by two fine-tuned generations. All experiments were run on an NVIDIA A100 $40$GB GPU. Embedding and preprocessing consumed $\sim100$ GPU hours; hyperparameter tuning and cross-validation added $\sim50$ GPU hours. More details are provided in our GitHub repository~\cite{gulati2024linkedin}.

\subsection{Training and evaluation scenarios}
We trained models on LLP vs FLP as a baseline, and introduced \ul{three attack and three adversarial training} scenarios using GPT3.5Ps, GPT4Ps, or both. Each model was evaluated on all four profile types (LLPs, FLPs, GPT3.5Ps, and GPT4Ps). Classifiers were trained using STE embeddings from RoBERTa, DeBERTa, ModernBERT, and Flair with XGBoost and CatBoost.

Models were evaluated using F1 score, false accept rate (FAR; fake → legitimate), and false reject rate (FRR; legitimate → fake). The effectiveness of the attack and countermeasure was measured by changes in FAR under adversarial conditions and after retraining. Calibration was assessed using reliability curves~\cite{pavlovic2025understanding}, which compare predicted probabilities to empirical frequencies; the diagonal indicating perfect calibration. Brier score~\cite{brier1950verification} was also computed as a scalar measure of calibration, where lower values indicate more reliable confidence estimates, critical for minimizing overconfident mis-classification of LLM-generated profiles.

\begin{table*}[t]
\centering
\small
\renewcommand{\arraystretch}{1.2}
\caption{Training and test splits across scenarios. Left: training data composition. Right: test sets for robustness evaluation.}
\begin{tabular}{
p{2.8cm}
>{\centering\arraybackslash}p{0.7cm}
>{\centering\arraybackslash}p{0.7cm}
>{\centering\arraybackslash}p{0.7cm}
>{\centering\arraybackslash}p{0.7cm}
@{\hspace{8pt}}
>{\columncolor[gray]{0.95}}p{2.5cm}
>{\columncolor[gray]{0.95}\centering\arraybackslash}p{0.7cm}
>{\columncolor[gray]{0.95}\centering\arraybackslash}p{0.7cm}
>{\columncolor[gray]{0.95}\centering\arraybackslash}p{0.7cm}
>{\columncolor[gray]{0.95}\centering\arraybackslash}p{0.7cm}
}
\hline
& \multicolumn{4}{c}{\textbf{Train set}} & & \multicolumn{4}{>{\columncolor[gray]{0.95}}c}{\textbf{Test/Attack set}} \\
\textbf{Train scenario} & 
\rotatebox{90}{LLPs} & 
\rotatebox{90}{FLPs} & 
\rotatebox{90}{GPT3.5Ps} & 
\rotatebox{90}{GPT4Ps} & 
\textbf{Test scenario} &
\rotatebox{90}{LLPs} & 
\rotatebox{90}{FLPs} & 
\rotatebox{90}{GPT3.5Ps} & 
\rotatebox{90}{GPT4Ps} \\
\hline
Baseline & $1260$ & $420$ & - & - & Baseline & $540$ & $180$ & - & - \\
GPT3.5 Retrain & $1260$ & $420$ & $840$ & - & GPT3.5 Attack & $540$ & $180$ & $360$ & - \\
GPT4 Retrain & $1260$ & $420$ & - & $420$ & GPT4 Attack & $540$ & $180$ & - & $180$ \\
GPT3.5+4 Retrain & $1260$ & $420$ & $840$ & $420$ & Combined Attack & $540$ & $180$ & $360$ & $180$ \\
\hline
\end{tabular}
\label{tab:dataset_splits}
\end{table*}

\subsection{Human and GPT-4 evaluation}
We benchmarked human and GPT-4 detectors on the same inputs: Name, Location, Education, Experience, Skills, Connections, Followers, Summary, and derived statistics. GPT-4 was tested on $360$ profiles ($180$ real, $180$ fake) using the OpenAI API with zero-shot (single input) and few-shot ($3$ random labeled examples) prompting. Profiles labeled fake included both FLPs and LLM-generated profiles. More details are included in the GitHub repository~\cite{gulati2024linkedin}.

Thirty human evaluators classified $15$ profiles each ($5$ LLPs, $5$ FLPs, $5$ LLM-generated) through a structured web portal~\cite{fake_profile_survey}. For each profile description presented, they selected one of three labels: legitimate, manual fake, or LLM fake. We collected a total of $450$ responses ($15$ profiles × $30$ participants). Results were binarized (fake vs legitimate) to enable comparison with classifier and GPT-4 performance.

\section{Results and discussion}
\label{sec:results}
\subsection{Feature and classifier selection}
\label{sec:pca_results}
\begin{figure}[htbp]
    \centering
    \begin{minipage}[b]{0.48\textwidth}
        \centering
        \includegraphics[width=\textwidth,height=0.7\textwidth]{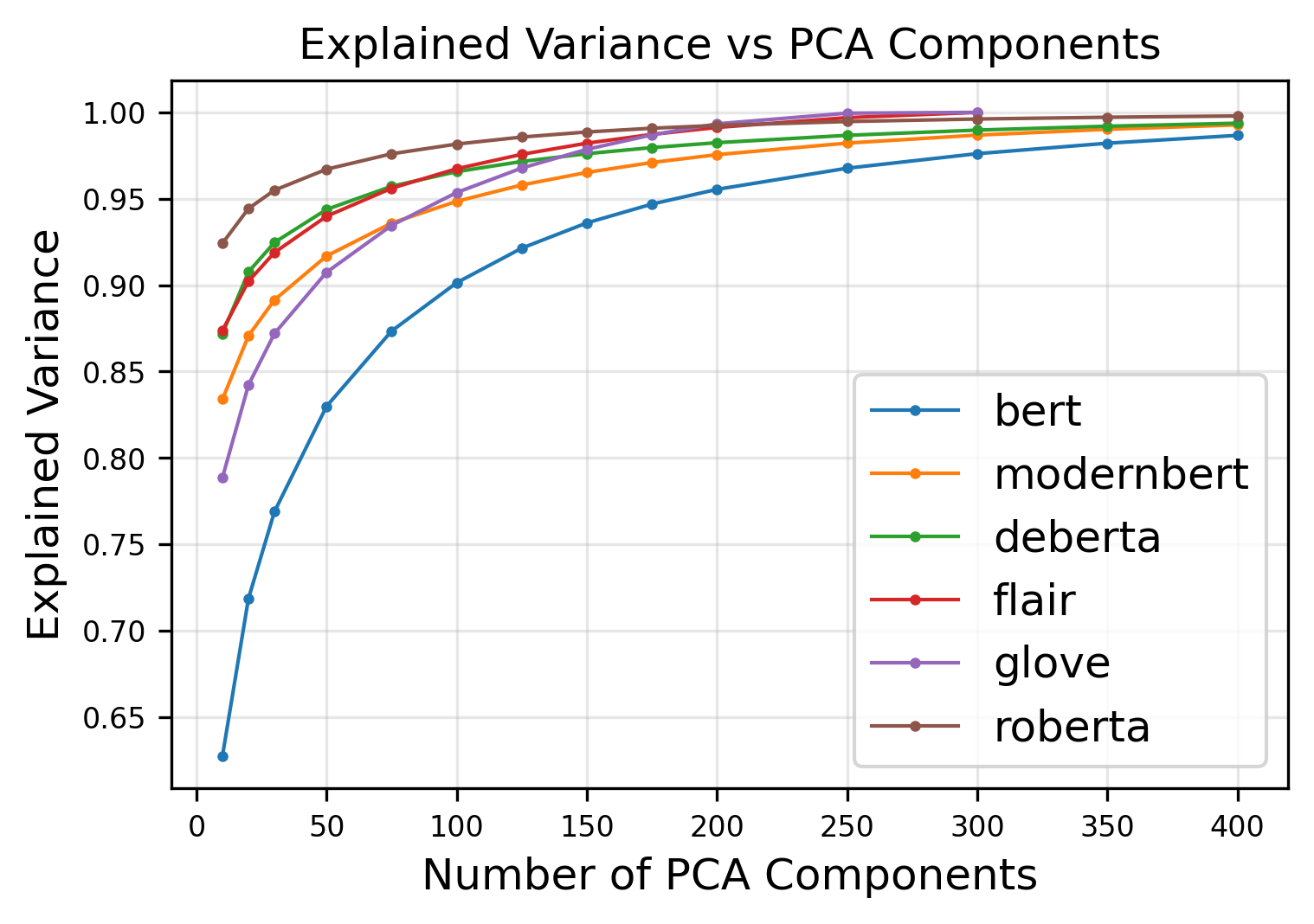}
    \end{minipage}
    \hfill
    \begin{minipage}[b]{0.48\textwidth}
        \centering
        \includegraphics[width=\textwidth,height=0.7\textwidth]{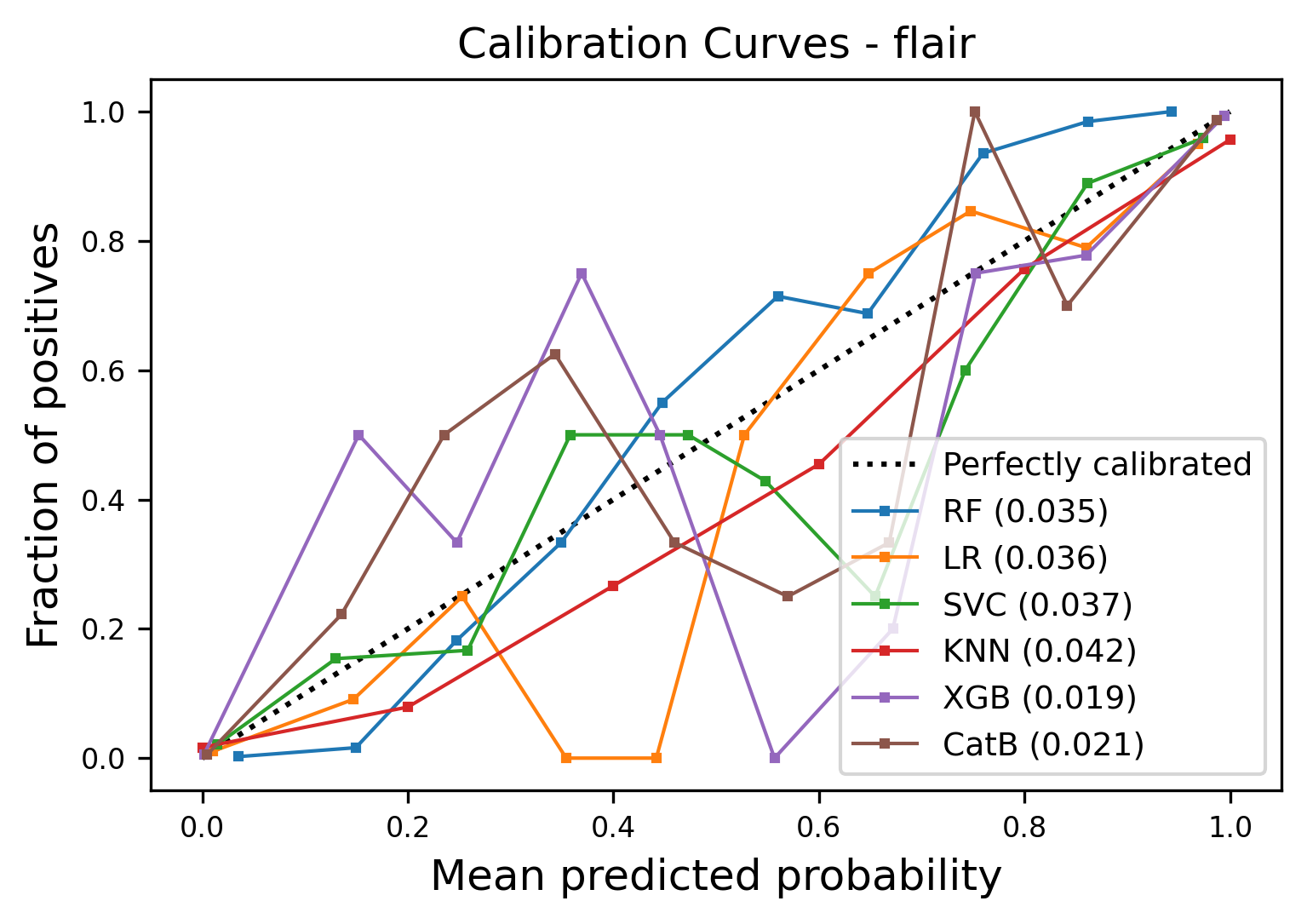}
    \end{minipage}
    \caption{(Left) PCA variance curves highlight that Flair and RoBERTa achieve faster variance saturation, suggesting higher intrinsic dimensional efficiency compared to BERT and GloVe. (Right) Flair-based calibration curves show boosting classifiers (XGBoost, CatBoost) align most closely with ideal calibration, as reflected in their low Brier scores; other models exhibit under- or overconfidence, particularly in mid-range probabilities.}
    \label{fig:combined}
\end{figure}

PCA analysis (Figure~\ref{fig:combined}, Left) shows that the top $150$ components captured $93.6-98.9$\% of variance across encoders, with negligible gains beyond this point. Consequently, we fixed $150$ components for RoBERTa, DeBERTa, ModernBERT, and Flair.
 
\begin{figure*}[h]
    \centering
    \begin{subfigure}{\textwidth}
        \centering
        \includegraphics[width=4in, height=0.9in]{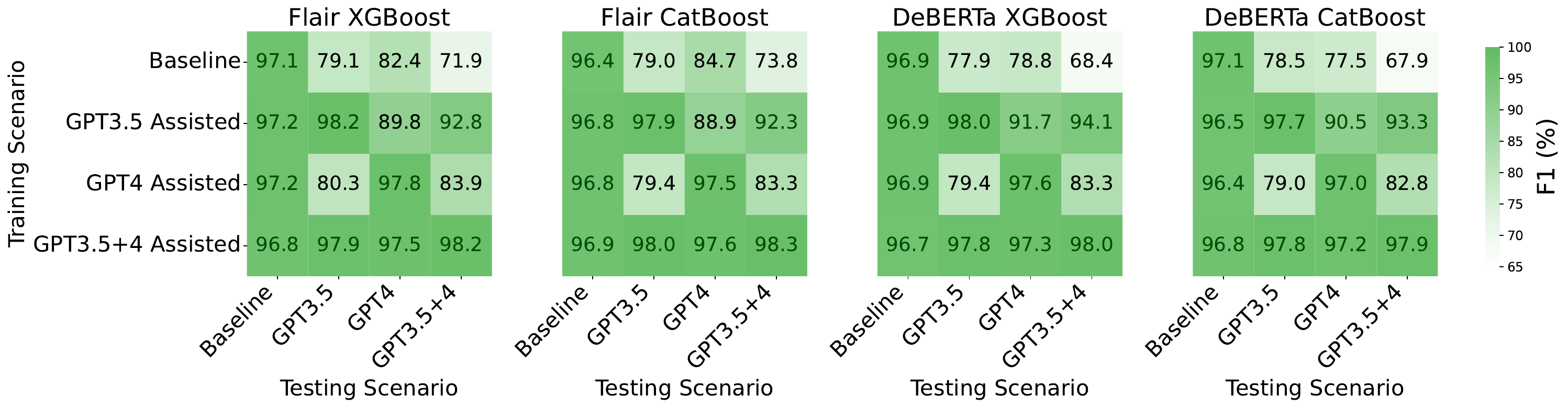}
        \caption{}
        \label{fig:f1_heatmap}
    \end{subfigure}
    
    \begin{subfigure}{\textwidth}
        \centering
        \includegraphics[width=4in, height=0.9in]{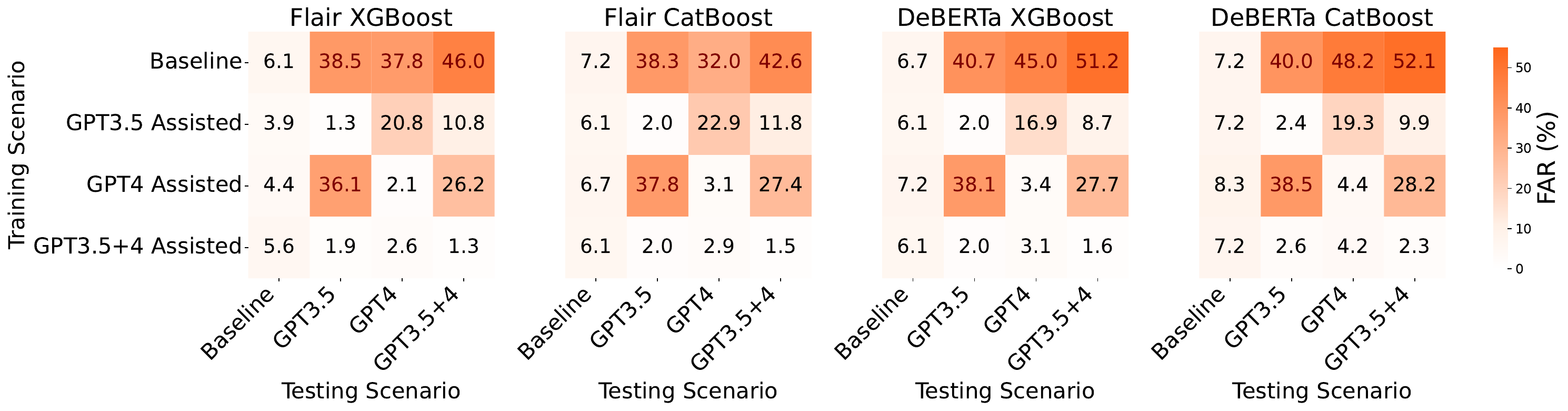}
        \caption{}
        \label{fig:far_heatmap}
    \end{subfigure}
    
    \caption{Performance of STE-based models with Flair and DeBERTa embeddings across different training and testing scenarios. (a) F1 show performance degradation under GPT-generated profile attacks, particularly in the baseline setting. (b) False Accept Rates (FAR) highlight model vulnerability to GPT3.5 and GPT4 profiles, with FARs exceeding 50\% in the worst case. Adversarial training—especially with combined GPT3.5+4P data—restores detection performance, yielding consistently high F1 scores and low FARs. Flair and DeBERTa results are shown; all classifiers and embeddings are included in the full evaluation (visit GitHub~\cite{gulati2024linkedin}).
   }
    \label{fig:combined_heatmaps}
\end{figure*}
 
Classifier calibration using Flair embeddings (Figure~\ref{fig:combined}, Right) showed that boosting models—CatBoost ($0.021$) and XGBoost ($0.019$)—achieved the lowest Brier scores, outperforming Logistic Regression ($0.036$), Random Forest ($0.035$), and KNN ($0.042$). XGBoost was slightly overconfident in the mid-range ($0.5–0.7$), while CatBoost maintained better reliability in high-confidence regions ($0.7–1.0$), which is critical for minimizing false accept rates. These calibration properties motivated the selection of CatBoost and XGBoost as our primary classifiers for further analysis \cite{kasneci2024enriching}.

\subsection{Performance under baseline and adversarial scenarios}
Figure~\ref{fig:combined_heatmaps} summarizes model performance across training and testing scenarios. 

\ul{Baseline performance:} Our STE-based models using Flair and DeBERTa embeddings outperformed prior work by \cite{Ayoobi_2023}, achieving F1 scores of $96.39\%–97.08$\%, compared to their reported $87.78\%–94.28$\% (STE) and $95.00\%–96.33$\% (SSTE).

\ul{Vulnerability to GPT-generated profiles:} On GPT3.5+4P attacks, F1 scores dropped to $67.88\%–73.82$\%, and FARs rose to $52.1$\% (DeBERTa+CatBoost), indicating over half of sophisticated fake profiles were misclassified as legitimate. This degradation aligns with high textual similarity between GPT-generated and real profiles (mean: $88.9$\%, range: $64.2\%–99.4$\%).

\ul{Adversarial training:} GPT3.5-assisted training improved F1 to $97.83\%–98.15\%$ and cut FARs on GPT3.5Ps to as low as $1.3\%$, but remained vulnerable to GPT4Ps (FARs: $16.9\%–19.3\%$). GPT4-assisted training reversed this—F1 up to $97.84$\% and FARs on GPT4Ps down to $\sim2$\%, but showed limited generalization to GPT3.5Ps. In contrast, training on the combined GPT3.5+4P dataset yielded strong generalization across all attacks, with FARs between $1.34$\% and $2.6$\% and F1 scores consistently above $97.5$\%. Flair+XGBoost achieved the best overall performance (F1 = $98.2$\%, FAR = $1.34$\% on combined attacks). Across all adversarial training settings, FRRs remained stable ($1.48$\%–$2.41$\%), confirming no significant compromise on correctly classifying legitimate profiles.

\subsection{LLM and human benchmarking results}
\label{sec:human_benchmark_results}
\begin{figure}[h]
    \centering
    \includegraphics[width=3.66in, height=1in]{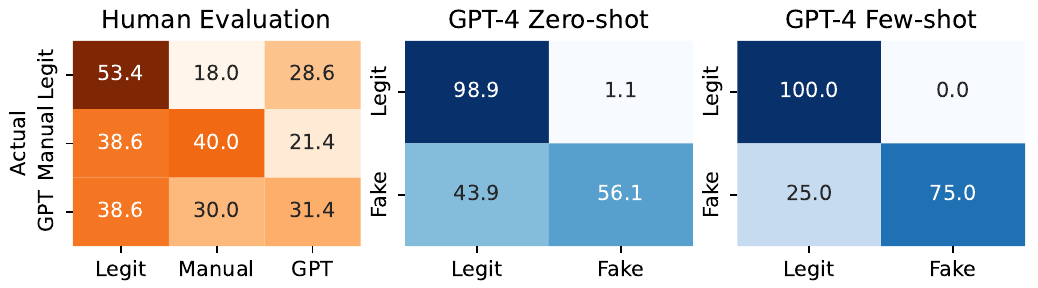} 
    \caption{
Confusion matrices comparing human and GPT-4 performance on LinkedIn fake profile detection. 
Left: Human annotators ($n = 30$) show moderate accuracy on legitimate profiles but perform poorly on fake ones, particularly GPT-generated cases. 
Middle: GPT-4 (zero-shot) detects legitimate profiles with high accuracy ($98.9\%$) but misclassifies nearly half of fake profiles. 
Right: Few-shot prompting improves GPT-4's fake detection to $75.0\%$ while maintaining $100\%$ accuracy on legitimate profiles.}
    \label{fig:combined-benchmark}
    \vspace{-0.4in}
\end{figure}

Figure~\ref{fig:combined-benchmark} compares the performance of human evaluators and GPT-4 on the task of LinkedIn fake profile detection. Human evaluators showed limited effectiveness, particularly on GPT-generated profiles, with a three-class accuracy of only $31.4\%$ on this category. Aggregated into a binary classification task, their F1 score was $58.9\%$, with a false accept rate (FAR) of $38.7\%$ and a false reject rate (FRR) of $46.6\%$, indicating considerable confusion between real and fake profiles. GPT-4 performed better overall. In the zero-shot setting, it reached an F1score of $71.3\%$ but misclassified $43.9\%$ of fake profiles. With few-shot prompting, its performance improved substantially, achieving perfect accuracy on legitimate profiles and reducing the FAR on fakes to $25.0\%$, raising the F1 score to $85.7\%$. However, both approaches fall short of our adversarially trained models, which consistently achieve F1 scores above $97.5\%$ and FARs below $2\%$ across all LLM-generated profile scenarios.

\subsection{Model robustness indicator}
To assess the link between model calibration and robustness to adversarial inputs, we computed Pearson correlations between Brier scores~\cite{brier1950verification} and false accept rates (FAR) across encoders and classifiers. Results show a consistent positive correlation: models with lower Brier scores—indicating better-calibrated confidence estimates—tended to exhibit lower FARs against LLM-generated fake profiles. DeBERTa embeddings showed particularly strong correlations (Pearson $r > 0.96$, $p < 0.001$), with CatBoost yielding the highest observed coefficients ($r \approx 0.978$–$0.979$). Flair embeddings demonstrated slightly weaker but still significant correlations ($r = 0.809$–$0.972$, $p < 0.001$). Adversarial training improved both calibration and robustness. For instance, DeBERTa with CatBoost reduced its Brier score from $0.135$ to $0.063$ and its FAR from $52.15\%$ to $2.78\%$ after GPT3.5+4P-assisted training. These findings support prior evidence that well-calibrated models with sharper decision boundaries are more resistant to high-quality adversarial content~\cite{qin2020improving, emde2024towards}.

\subsection{Ablation study}
\ul{ Contributions of the text and numerical embedding (baseline and attack)} 
To assess the contribution of different feature modalities, we compared models trained using only STE embeddings ($150$ dimensions) versus only numerical features ($17$ dimensions). Under baseline conditions, both configurations performed comparably: STE achieved F1 scores of $95.19\%$ (CatBoost) and $96.23\%$ (XGBoost), while numerical features reached $95.56\%$ and $95.04\%$, respectively.

Under adversarial attack, STE-based models experienced substantial performance degradation. F1 scores fell to $67.1\%$–$77.17\%$ (GPT3.5Ps), $75.16\%$–$81.41\%$ (GPT4Ps), and $57.49\%$–$71.63\%$ (combined). In contrast, models using only numerical features exhibited greater robustness, with F1 scores consistently in the range of $78.67\%$–$80.63\%$ across attack types.

These results suggest that while textual features are effective against manually constructed fakes, they are more susceptible to LLM-generated attacks. Numerical features, although lower in dimensionality, appear to capture structural patterns that generalize more effectively in adversarial contexts. This complementarity underscores the benefit of combining both feature types for robust detection.

\ul{ Text vs. numerical features (post-adversarial training)}  
Feature types also responded differently to adversarial training. Using only STE embeddings, GPT4P-assisted training reduced FARs to $2.59\%$–$8.33\%$ (GPT3.5Ps) and $3.12\%$–$8.33\%$ (GPT4Ps), indicating improved generalization across LLM variants. 

In contrast, models using numerical features demonstrated asymmetric gains. For the full $167$-dimensional feature set, GPT4P-assisted training substantially reduced FARs for GPT4Ps but had a limited impact on GPT3.5Ps. For example, Flair+XGBoost: FAR dropped from $38.52\%$ to $36.11\%$. A similar trend was observed when using only numerical features: GPT3.5P-assisted training improved F1 scores from $78.67\%$–$79.21\%$ to $82.55\%$–$84.08\%$ on GPT3.5Ps, whereas GPT4P-assisted training led to a larger jump for GPT4Ps (up to $96.75\%$).

These findings suggest that textual features tend to generalize more effectively across model variants and adversarial scenarios, whereas numerical features seem to encode generation-specific artifacts. Moreover, the compact 17-dimensional numerical representation offers a lightweight alternative for detection in resource-constrained settings.

\section{Limitations and future research directions}
\label{sec:future_dir}
While our approach achieves low FARs ($1.34\%$–$2.28\%$) through STE-based features, PCA, and adversarial training under both adversarial and non-adversarial environments, several limitations remain. First, the evaluation is restricted to English-language LinkedIn profiles and should be extended to languages other than English. The method’s applicability to other platforms or multilingual contexts has not been tested. Second, embedding extraction relies on a fixed set of LLM-based encoders (e.g., DeBERTa, RoBERTa), which may affect stability as model architectures evolve. Third, both the creation of fake profiles for generating attack vectors and
adversarial training used models from the same LLM family (OpenAI GPT). The system needs to be evaluated against a wide variety of advanced LLMs. Fourth, the current experiments should be expanded to include legitimate profiles created by legitimate people who utilized LLMs for creating and polishing their profiles. 

\section{Conclusion}
\label{sec:conclusion}
Existing LinkedIn fake profile detectors perform well on manually created profiles (F1 $>$ $95$\%) but fail on GPT3.5 and GPT4-generated ones, with F1 dropping to 67.88\% and false accept rates (FAR) exceeding $52$\%. Human annotators (F1 = $58.9$\%) and general-purpose LLMs (F1 = $85.7$\%) also underperform in this setting. Targeted adversarial training using GPT-generated profiles restored F1 to $98.2$\% and reduced FAR to $1.34$\%, with minimal impact on legitimate profile rejection (FRR $<$ $2.5$\%). Flair embeddings with XGBoost gave the most consistent results. Ablation experiments revealed that textual features degrade sharply under attack, while numerical features remain more robust. Their combination yields better generalization across model variants and input conditions. These findings support the need for task-specific retraining to maintain robustness against high-quality synthetic profiles generated with the help of LLMs.

\bibliographystyle{IEEEtran}
\bibliography{asonam}

\end{document}